\documentclass{llncs}
\begin{document}

\newcommand\Z{{\mathbb Z}}

\newcommand\U{{\mathbf U}}

\newcommand\J{{\mathcal J}}
\newcommand\K{{\mathcal K}}
\newcommand\E{{\mathcal E}}
\newcommand\D{{\mathcal D}}
\newcommand\A{{\mathcal A}}
\newcommand\B{{\mathcal B}}
\newcommand\SE{\mathcal{SE}}

\title{Concrete Security Against Adversaries with Quantum
  Superposition Access to Encryption and Decryption Oracles}

\author{Shahram Mossayebi and R\"udiger Schack}
\institute{Royal Holloway, University of London, Egham, Surrey TW20 0EX, UK}

\maketitle

\begin{abstract}
In 2013, Boneh and Zhandry introduced the notion of indistinguishability (IND)
in chosen plaintext (CPA) and chosen ciphertext (CCA) attacks by a quantum
adversary which is given superposition access to an oracle for encryption and
decryption queries but is restricted to classical queries in the challenge
phase. In this paper we define IND-CPA and IND-CCA notions for symmetric
encryption schemes where the adversary has full quantum superposition access to
the oracle, and give constructions that achieve these security notions. Our
results are formulated in the concrete security framework.
\end{abstract}

\section{Introduction}

Even though scalable quantum computers cannot be built using current
technology, the fact that they may become possible in the future has an impact
on present-day information security. The relatively recent field of
post-quantum cryptography \cite{Bernstein2009}
 therefore studies classical schemes that remain secure if the
adversary possesses a quantum computer. (Here and throughout the paper,
``classical'' is taken to mean ``non-quantum''.) Since the subject of
post-quantum cryptography is the security of present classical technology
against future quantum attacks, the usual assumption is that only the adversary
possesses quantum capabilities. This means that all communication between the
adversary and the legitimate parties, in particular oracle access, is assumed
to be classical.

In 2013, Boneh and Zhandry \cite{Boneh2013b} went beyond this paradigm by
introducing a security model in which the adversary is given quantum
superposition access to encryption and decryption oracles. Their work is of
considerable conceptual interest. In addition, it may become practically
relevant in a future technological landscape where some cryptographic protocols
are implemented on quantum computers.

Boneh and Zhandry define a notion of indistinguishability (IND) in chosen
plaintext (CPA) and chosen ciphertext (CCA) attacks through a game with two
phases. In the first phase (the query phase) the adversary is given
unrestricted superposition oracle access. In the second phase (the challenge
phase) the adversary is only allowed to make classical queries. 

Why is there a restriction in the challenge phase? In their paper, Boneh and
Zhandry show that some restriction is necessary. The reason is, effectively,
that a quantum computer can easily distinguish between the encryption of a
single message and the encryption of an equal superposition of all possible
messages, independently of the details of the encryption. The restriction to
classical challenge queries prevents the adversary to exploit this fact and
allows Boneh and Zhandry to prove that their security notion is achievable. 
It is worth pointing out that even in standard (classical) indistinguishability
notions it is necessary to restrict the class of allowed challenge queries, to
prevent the adversary from winning the game trivially by exploiting information
about, e.g., message length.

There is a sense, however, in which a restriction to classical challenge
queries seems too strong. Considering quantum oracle access makes sense only in
view of a future technological environment in which the legitimate parties use
quantum computers. In such an environment, it is likely that the encryption
schemes considered here will form part of a wider quantum communication
infrastructure. Without a precise specification of the nature of this
infrastructure, one should not rule out a priori the possibility that an
adversary might benefit from the ability to distinguish between superpositions
of messages or ciphertexts. 

In this paper we introduce an achievable security notion where the adversary
has full superposition access to an oracle. We define an indistinguishability
(IND) notion in both chosen plaintext (CPA) and chosen ciphertext (CCA)
attacks. Our IND notion, which we call ``real or permutation'' (RoP), is
equivalent to standard IND notions in the case of classical oracle queries, but
is immune to the attack discovered by Boneh and Zhandry in the case of quantum
superposition queries. It also falls outside the classification of security
notions recently given by Gagliardoni, H\"ulsing and Schaffner
\cite{Gagliardoni2015}. It would be worthwhile to study the
relationship between our CPA notion and the qIND-qCPA notion defined in
\cite{Gagliardoni2015}. Whereas we consider direct quantum-mechanical
interaction between adversary and oracle, qIND-qCPA requires the adversary to
submit classical descriptions of its quantum queries.

In a RoP experiment, the adversary is given access to one of two encryption
oracles. One oracle simply encrypts challenge messages chosen by the adversary,
whereas the other oracle applies a random permutation to the message and then
encrypts it. The adversary's goal is to distinguish between the two cases. For
classical queries, applying a random permutation to a message is equivalent to
replacing the message by a random string. For classical queries, the RoP
security notion is therefore equivalent to the ``real or random'' notion
defined in \cite{Bellare1997}. In the quantum case, the RoP
notion allows for arbitrary quantum superposition queries in all phases of the
experiment. 

The paper is organized as follows. In Section~\ref{sec:model} we describe our
security model, for which we adopt the concrete-security paradigm
\cite{Bellare1997}.  Section~\ref{sec:qprf} discusses quantum pseudorandom
functions (QPRF) and their existence from the concrete-security standpoint. In
Section~\ref{sec:cpa} we define our real or permutation IND-CPA notion and show
that it is achieved by a slight modification of a standard construction.
The CCA case is the topic of Section~\ref{sec:cca}. The
proof that the RoP IND-CCA notion is achievable is the main contribution of
this paper. The proof contains new ideas and substantial input from quantum
information theory.

\newpage

\section{Security Model}  \label{sec:model}

In this paper we address the security of symmetric encryption schemes against
quantum adversaries \cite{Hallgren2011,Unruh2012}. We adopt the usual
definition of a symmetric encryption scheme as a triple $\SE=(\K,\E,\D)$
consisting of a (randomized) key generation algorithm $\K$, a (randomized) encryption
algorithm that takes a plaintext $M$ and a key $K$ and returns a ciphertext
$C=\E_K(M)$, and a decryption algorithm satisfying $\D_K(\E_K(M))=M$ for all
messages $M$.  We assume that $\E$ is randomized, i.e., its output depends on a
string $r$ which is chosen randomly each time $\E$ is invoked.

We will consider both chosen plaintext attacks (CPA), where an adversary is given
access to an encryption oracle, and chosen ciphertext attacks (CCA), where the
adversary in addition has access to a decryption oracle. We will assume that both $\E$
and $\D$ are implemented on a quantum computer \cite{Nielsen2000}, 
and that the adversary is given
superposition access to the corresponding oracles. By this we mean, roughly,
that the adversary can make encryption queries consisting of quantum
superpositions of messages, to which the oracle responds with a corresponding
superposition of ciphertexts, and similarly for decryption queries in the CCA
case. 

Formally we define a superposition query as follows. For any function
$f:\{0,1\}^n\to\{0,1\}^m$ we define a unitary transformation, $\mathbf{U}_f$,
on a $(n+m)$-qubit register by 
\begin{equation}
\label{eq:GeneralQuantumQuery}
	\mathbf{U}_{f} | x, y \rangle = |x, y \oplus f(x) \rangle \;,
\end{equation} 
where $x$ is an $n$-bit string, $y$ is an $m$-bit string, and 
\begin{equation}
|x,y\rangle = |x\rangle|y\rangle = |x\rangle\otimes|y\rangle \;,
\end{equation}
the {\it computational basis states}, form an orthonormal basis of
$2^{n+m}$-dimensional complex Hilbert space. Equation
(\ref{eq:GeneralQuantumQuery}) defines the action of $\mathbf{U}_f$ for
arbitrary quantum states on the quantum register, including superposition
states and including the case that the register is entangled with some other
quantum register.

In this scenario, an encryption query consists in the application of the
unitary $\U_{\E_K}$ to a quantum register under the control of the
adversary, and similarly for a decryption query. We place no restrictions
on the initial state of the adversary's register. If the initial state is of
the superposition form $|\psi\rangle=\sum_Mc_M|M,0\rangle$ with arbitrary
complex coefficients $c_M$, the result of the query is
\begin{equation}  \label{eq:superpositionQuery}
\U_{\E_K}|\psi\rangle = \sum_M c_M|M,\E_K(M)\rangle\;.
\end{equation}

The resources required to apply the unitary $\U_{\E_K}$ to a quantum register
are independent of the initial state $|\psi\rangle$ of the register. Applying a
unitary can be thought of as a single physical operation, for which the number
of terms in the superposition state $|\psi\rangle$ is irrelevant. Since the
encryption oracle does not ``know'' whether it acts on a superposition or on a
single basis state, we will assume that the random string $r$ required for the
randomized encryption is chosen exactly once every time $\U_{\E_K}$ is
applied. This means that $r$ is the same for all terms in the sum in
Eq.~(\ref{eq:superpositionQuery}).

In the most general definition, a \textit{quantum adversary\/} $\A$ is a
quantum algorithm that runs on a quantum computer. In this paper, we will
assume that $\A$ takes an (optional) bit string $i$ as input, has access to one
or more oracles $f_1,f_2,\ldots$ and eventually halts, outputting a bit string
$o$. We will denote this process by
\begin{equation}
o \leftarrow \A^{f_1,f_2,\ldots}(i) \;.
\end{equation}
We will assume that $\A$ maintains a quantum register $Q_\A$ for the purpose of
making oracle queries, a quantum register $S_\A$ for doing quantum computations
and for storing its internal state between invocations, and a register $R_\A$
for classical input and output. Passing an argument $i$ to $\A$ is done by
placing $i$ into the register $R_\A$. One could model $R_\A$ as a quantum
register, but in practice one would expect quantum algorithms to have classical
as well as quantum parts. Whenever $\A$ makes a query to an oracle $f$, the
unitary operation $U_f$ defined in Eq.~(\ref{eq:GeneralQuantumQuery}) is
applied to the register $Q_\A$.  Since it is implicit in
Eq.~(\ref{eq:GeneralQuantumQuery}) that $U_f$ is applied to $n+m$ qubits, we
will assume that $\A$ puts the parameters $n$ and $m$ in the register $R_\A$ if there is
any ambiguity.

In this paper we adopt the concrete security framework \cite{Bellare1997}. 
Instead of focusing on polynomial algorithms in an asymptotic
sense, concrete security concerns bounds on an adversary's success probability
as a function of the actual resources available to the adversary.

The most relevant resources for the purposes of this paper are the running time
of a quantum adversary $\A$, and the number of oracle queries made by $\A$. We
define the running time as the time, in seconds, that elapses until $\A$ writes
its final output and halts, including any initialization steps. There exist a
number of further potentially important resource parameters, such as memory size, or the
number of qubits required by $\A$, but since these do not play any explicit
role in the reduction arguments given below, we will not discuss them here.

Our reduction arguments can be read in a very ``concrete'' way, e.g., ``if
there exists a specific quantum adversary that, in $10^4$ seconds and using
$10^9$ oracle queries achieves an advantage of $2\times10^{-2}$ in an attack on
scheme $X$, then one can construct another quantum adversary that, also in
$10^4$ seconds and using $10^9$ oracles queries, achieves an advantage of at
least $10^{-2}$ in an attack on scheme $Y$.'' Working within the
concrete-security paradigm and measuring running time in seconds rather than as
the number of, say, gate operations has the clear advantage that it leads to
definitions and theorems which are independent of any particular computing or
quantum computing model.

\section{Quantum pseudorandom functions}  \label{sec:qprf}

In a concrete security framework, a quantum pseudorandom function, or QPRF, is
simply a family of functions. What turns a family of functions into a QPRF is a pair of 
experiments that defines an adversary's {\it QPRF advantage}.

So let $F: \mathcal{K} \times \mathcal{X} \rightarrow \mathcal{Y}$ be a
function family identified by the set $\mathcal{K}$. Consider two oracles: one
formed by an instance $F_K=F(K,\cdot)$ for a random $K\in\K$, the other given
by a function $f$ chosen randomly from the set
$\mathsf{Func}(\mathcal{X},\mathcal{Y})$ of functions from $\mathcal{X}$ to
$\mathcal{Y}$.  The QPRF advantage is a measure of the adversary's ability to
distinguish an instance drawn from the function family from a function chosen
at random from $\mathsf{Func}(\mathcal{X},\mathcal{Y})$.

\begin{figure}[tp]
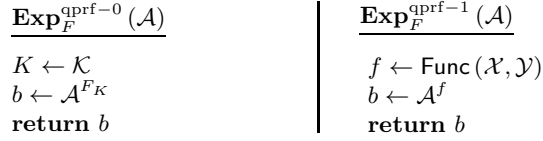

	\centering
		\begin{tabular}{l|l}
			\begin{minipage}{0.33\textwidth}

				\underline{$\textbf{Exp}_{F}^{{\rm qprf}-0}\left(\mathcal{A}\right)$}
				\vspace*{0.75em}
				\\
				$K \leftarrow \mathcal{K}$ \\
				$b \leftarrow \mathcal{A}^{F_K}$ \\
				\textbf{return} $b$

			\end{minipage}
			&
			\begin{minipage}{0.33\textwidth}

				\hspace*{5mm}\underline{$\textbf{Exp}_{F}^{{\rm qprf}-1}\left(\mathcal{A}\right)$}
				\vspace*{0.75em}
				\\
				\hspace*{5mm} $f \leftarrow \mathsf{Func}\left(\mathcal{X}, \mathcal{Y}\right)$ \\
				\hspace*{5mm} $b \leftarrow \mathcal{A}^{f}$ \\
				\hspace*{5mm} \textbf{return} $b$

			\end{minipage} 
		\end{tabular}
	\caption{The two experiments defining a QPRF.}
	\label{fig:QPRF}
\end{figure}

\begin{definition}[\textbf{QPRF}]
\label{def:QPRF}
Define experiments $\textbf{Exp}_{F}^{{\rm qprf}-0}$ and
$\textbf{Exp}_{F}^{{\rm qprf}-1}$ as in Figure~\ref{fig:QPRF}.  In both
experiments, the quantum adversary $\A$ is given access to an oracle and eventually
outputs a bit, $b$. In $\textbf{Exp}_{F}^{{\rm qprf}-0}$, a key $K\in\K$ in
chosen at random and the adversary's oracle queries are answered by
applying the unitary operator $\U_{F_K}$ to the adversary's register $Q_\A$.
In $\textbf{Exp}_{F}^{{\rm qprf}-1}$, a function
$f\in\mathsf{Func}(\mathcal{X},\mathcal{Y})$ is chosen at random and the adversary's
oracle queries are answered by similarly applying the unitary operator $\U_{f}$.
The QPRF advantage of $\mathcal{A}$ is defined as
\[
\textbf{Adv}_{F}^{{\rm qprf}} \left(\mathcal{A}\right) 
= \rm{Pr} \left[ \textbf{Exp}_{F}^{{\rm qprf}-1} \left(\mathcal{A}\right) 
= 1 \right] - \rm{Pr} \left[ \textbf{Exp}_{F}^{{\rm qprf}-0}  \left(\mathcal{A}\right) 
= 1 \right]\;.
\]
\end{definition}

In the next two sections, we will analyze the security of encryption schemes
which are based on a QPRF. For these schemes to be secure, we need to assume
that there exists a function family $F$ such that its QPRF-advantage is
extremely small for any quantum adversary using resources that are available
now or might become available in the foreseeable future. Such function families
are widely believed to exist in the form of standard block ciphers, for
instance AES-256. The best currently known quantum attack against AES-256 uses
Grover's search algorithm \cite{Grover1996,Nielsen2000} and requires of the
order of $2^{128}$ queries to find the encryption key with high probability. The security of
the schemes discussed below depends on the heuristic assumption that AES-256 or
similar block ciphers cannot be broken by a quantum computer using realistic
resources.

\section{Quantum Superposition Chosen Plaintext Attack}   \label{sec:cpa}

To define indistinguishability in a CPA attack for a symmetric encryption
scheme $\SE$, we introduce a pair of experiments as in
Figure~\ref{fig:RoPqsCPA}.  We call them real or permutation, or RoP,
experiments, in analogy to the real or random notion defined in
\cite{Bellare1997}. In each experiment, the quantum adversary is given
superposition access to an encryption oracle.  The encryption oracle responds
to each encryption query by applying a unitary transformation to the
adversary's quantum register $Q_\A$.  The transformation depends on the bit
$b$.  If $b=1$, the transformation is given by
\begin{equation}
 \left|m,x\right\rangle \longrightarrow  \left|m,x \oplus \E_K\left(m\right) \right\rangle\;, 
\end{equation}
and if $b=0$, it is
\begin{equation}
\left|m,x\right\rangle \longrightarrow
\left|m,x \oplus \E_K \left(\Pi \left(m\right)\right) \right\rangle \;,
\end{equation}
where $\Pi$ is a permutation chosen uniformly at random. This means that in the
case $b=0$, before the encryption a random permutation is applied to each term
in the superposition of plaintexts. The goal of the quantum
adversary is to distinguish between the two experiments.

\begin{figure}[tp]
	\centering
	\begin{tabular}{l}
		\underline{$\textbf{Exp}_{\mathcal{S}\mathcal{E}}^{rop-qscpa-b}\left(\mathcal{A}\right)$}
		\vspace*{0.75em}
		\\				
		\hspace*{5mm} $K \leftarrow \!\!{\scriptstyle \$}\, \mathcal{K}$ \\
		\hspace*{5mm} $b' \leftarrow \!\!{\scriptstyle \$}\, \mathcal{A}^{\mathsf{RoP}_{Q_{\mathcal{A}}}\left(\right)}$\\
		\hspace*{5mm} \textbf{return} $b'$
		\\
		\vspace*{1em}
		\\
		\underline{$\mathsf{RoP}_{Q_{\mathcal{A}}}\left(\right)$}
		\vspace*{0.75em}
		\\		
		\hspace*{5mm} \textbf{if} $b=1$ \textbf{then}\\
		\hspace*{10mm} Apply $\mathbf{U}_{\mathcal{E}_K \left(\cdot\right)}$ to $Q_{\mathcal{A}}$\\
		\hspace*{5mm} \textbf{else}\\
		\hspace*{10mm} \indent $\Pi \leftarrow \!\!{\scriptstyle \$}\, \mathsf{Perm} \left(n\right)$\\
		\hspace*{10mm} Apply $\mathbf{U}_{\Pi \left(\cdot\right)}$ to $Q_{\mathcal{A}}$\\
		\hspace*{10mm} Apply $\mathbf{U}_{\mathcal{E}_K \left(\cdot\right)}$ to $Q_{\mathcal{A}}$\\
		\hspace*{5mm} \textbf{end if}\\
		\hspace*{5mm} \textbf{return}\\
	\end{tabular}
	\caption{The RoP-qsCPA confidentiality notion}
	\label{fig:RoPqsCPA}
\end{figure}

\begin{definition}[\textbf{RoP-qsCPA}]
	Let $\mathcal{S}\mathcal{E}=\left(\mathcal{K}, \mathcal{E},
        \mathcal{D}\right)$ be a symmetric encryption scheme.  Define
        experiment
        $\textbf{Exp}_{\mathcal{S}\mathcal{E}}^{rop-qscpa-b}\left(\mathcal{A}\right)$
        for a quantum adversary $\mathcal{A}$ and a bit $b$ as shown in
        Figure~\ref{fig:RoPqsCPA}.  In the experiment, the adversary
        $\mathcal{A}$ is given quantum superposition access to a
        real-or-permutation encryption oracle
        $\mathsf{RoP}_{Q_{\mathcal{A}}}\left(\right)$.  The encryption oracle
        responds to each query by applying a unitary transformation to the
        adversary's quantum register $Q_{\mathcal{A}}$.
	
	The adversary's goal is to output a bit $b'$ as its guess of the
        challenge bit $b$, and the experiment returns $b'$ as well.  The
        advantage of $\A$ is given by:
	\[
		\textbf{Adv}^{rop-qscpa}_{\mathcal{S}\mathcal{E}}\left(\mathcal{A}\right) = 
		{\rm Pr}\left[\textbf{Exp}_{\mathcal{S}\mathcal{E}}^{rop-qscpa-1}\left(\mathcal{A}\right) = 1 \right] - 
		{\rm Pr}\left[\textbf{Exp}_{\mathcal{S}\mathcal{E}}^{rop-qscpa-0}\left(\mathcal{A}\right) = 1 \right] \;.
	\]
	This advantage refers to a specific quantum adversary using resources
        as defined in Section~\ref{sec:model}.\qed
\end{definition}

We now show that the above RoP-qsCPA indistinguishability notion can be
achieved. To motivate our construction, we first show that 
the following standard construction (see, e.g., \cite{Katz2007}) is insecure
with respect to our notion :\\

\noindent{\bf Construction 1':}
Let $F$ be a QPRF. The following construction
defines a symmetric encryption scheme $\mathcal{S}\mathcal{E}=\left(\mathcal{E}, \mathcal{D}\right)$:
	\begin{eqnarray*}
		\mathcal{E}\left(K,m\right):\, && r \leftarrow \!\!{\scriptstyle \$}\,	\left\{0,1\right\}^*\\
		&& c \leftarrow F_K\left(r\right) \oplus m\\
		&& \textbf{\em output} \; \left( r, c\right)\\
		\mathcal{D}\left(K,r,c\right):\, && m \leftarrow F_K\left(r\right) \oplus c\\
		&& \textbf{\em output} \; \left( m\right)
	\end{eqnarray*}

Suppose the state ${\cal N}\sum_m|0,m\rangle$ is submitted to the encryption
oracle, where ${\cal N}$ is a normalization constant. In the ``real case''
($b=1$), the result is the state ${\cal N}\sum_m|m,m\oplus y\rangle$, where 
$y=F_{K}(r)$. In the ``permutation'' case ($b=0$), the result is the state
${\cal N}\sum_m|m,\Pi(m\oplus y)\rangle$. A Fourier transform 
followed by a measurement will distinguish these two states with probability
almost 1.

The problem is that the same randomness $r$ is used for all terms in the
superposition. The following modified construction overcomes this problem.\\

\noindent{\bf Construction 1:}
Let $F$ be a QPRF. The following construction
defines a symmetric encryption scheme $\mathcal{S}\mathcal{E}=\left(\mathcal{E}, \mathcal{D}\right)$:
	\begin{eqnarray*}
		\mathcal{E}\left(K,m\right):\, && r \leftarrow \!\!{\scriptstyle \$}\,	\left\{0,1\right\}^*\\
		&& s \leftarrow F_r\left(m\right)\\
		&& c \leftarrow F_K\left(s\right) \oplus m\\
		&& \textbf{\em output} \; \left( s, c\right)\\
		\mathcal{D}\left(K,s,c\right):\, && m \leftarrow F_K\left(s\right) \oplus c\\
		&& \textbf{\em output} \; \left( m\right)
	\end{eqnarray*}

To prove that the modified construction achieves our notion of RoP-qsCPA
security, we are going to provide a straightforward reduction proof. In 
the concrete security framework adopted here this means that, if the above
construction can be broken by a specific quantum adversary, the
reduction establishes the existence of a quantum adversary using similar
resources that breaks the underlying QPRF. But as we saw in
Section~\ref{sec:qprf}, a QPRF based on a suitably chosen block cipher is
currently thought to be secure against quantum attacks.

\begin{theorem}[RoP-qsCPA security is achievable]
\label{thm:qsCPA}
Let $\mathcal{A}$ be a quantum adversary attacking the encryption scheme
$\mathcal{S}\mathcal{E}$, based on a QPRF $F$ as in Construction~1, in the
RoP-qsCPA sense. Assume $\A$ makes at most $q$ queries to the encryption oracle
and has advantage
\[
\textbf{\em{Adv}}_{\mathcal{S}\mathcal{E}}^{rop-qscpa}\left(\mathcal{A}\right) \ge \epsilon \;.
\]
Then there exists a quantum adversary $\mathcal{B}$ attacking $F$, making at most $q$ queries
to the encryption oracle and having advantage
\[
\textbf{\em{Adv}}_F^{qprf}\left(\mathcal{B}\right) \ge \frac{\epsilon}{2(q+1)} \;.
\]
\end{theorem}

{\bf Proof}. To prove the theorem one (i) establishes the security of the
scheme when $F$ is replaced by a truly random function $f$. Then (ii) one shows
that, if the scheme is insecure when the QPRF $F$ is used, then there exists a
quantum adversary which can distinguish $F$ from a truly random function and
thus breaks $F$.

For part (i), assume that in Construction 1, the QPRF $F_K$ is replaced by a
random function $g$, and in the $j$-th invocation of the encryption oracle
($j=1,\ldots,q$), the QPRF $F_r$ is replaced by a random function $f_j$. Assume
that the length of $s$ is chosen so large that computing $f_j(m)$ for all $m$
and $j$ leads to collisions with exponentially small probability. For
simplicity we assume that the same QPRF is used throughout, if necessary by
padding keys and/or arguments. No collisions means that for all $m_1,m_2$ in
the message space and for all $i,j$, $f_i(m_1)=f_j(m_2)\Rightarrow i=j \mbox{
  and } m_1=m_2$. Then $g(f_j(m))$ for all $m$ and $j$ are independent random
strings. Therefore the set $\{(f_j(m),g(f_j(m))\oplus m)\}$ is
information-theoretically indistinguishable from the set
$\{(f_j(m),g(f_j(m))\oplus\Pi(m))\}$. It follows that having access to superpositions of
states from the one or the other set cannot give rise to a positive advantage.

The above argument relies on the fact that information-theoretic notions carry
over to the quantum case. This is the only quantum argument needed in this
proof. Part (ii) of the proof proceeds by a standard hybrid argument that
assumes only classical queries. It is therefore omitted here.

\section{Quantum Superposition Chosen Ciphertext Attack}  \label{sec:cca}

In a CCA attack against a symmetric encryption scheme $\SE$ the quantum
adversary is given, in addition to an encryption oracle as in the CPA case,
superposition access to a decryption oracle. To define indistinguishability in
this case, we introduce the pair of experiments in
Figure~\ref{fig:RoPqsCCA}. The decryption oracle responds to each decryption
query by applying the following unitary transformation to the quantum
adversary's register $Q_\A$:
\begin{equation}
|x,c\rangle \longrightarrow |x\oplus \D_K(c),c\rangle \;.
\end{equation}
To arrive at a meaningful definition, we have to exclude decryption queries
consisting in the results of encryption queries. 

\begin{figure}[tp]
\centering
\begin{tabular}{l}
	\underline{$\textbf{Exp}_{\mathcal{S}\mathcal{E}}^{rop-qscca-b}\left(\mathcal{A}\right)$}
	\vspace*{0.75em}
	\\				
	\hspace*{5mm} $K \leftarrow \mathcal{K}$ \\
	\hspace*{5mm} $b' \leftarrow \mathcal{A}^{\mathsf{RoP}_{Q_{\mathcal{A}}}\left(\right), \mathsf{Dec}_{Q_{\mathcal{A}}}\left(\right)}$\\
	\hspace*{5mm} \textbf{return} $b'$
	\\
        \vspace*{1em}
	\\
	\underline{$\mathsf{RoP}_{Q_{\mathcal{A}}}\left(\right)$}
	\vspace*{0.75em}
	\\		
	\hspace*{5mm} \textbf{if} $b=1$ \textbf{then}\\
	\vspace*{0.75em}
	\hspace*{10mm} Apply $\mathbf{U}_{\mathcal{E}_K \left(\cdot\right)}$ to $Q_{\mathcal{A}}$\\
	\hspace*{5mm} \textbf{else}\\
	\hspace*{10mm} \indent $\Pi \leftarrow \!\!{\scriptstyle \$}\, \mathsf{Perm} \left(n\right)$\\
	\hspace*{10mm} Apply $\mathbf{U}_{\Pi \left(\cdot\right)}$ to $Q_{\mathcal{A}}$\\
	\hspace*{10mm} Apply $\mathbf{U}_{\mathcal{E}_K \left(\cdot\right)}$ to $Q_{\mathcal{A}}$\\
	\hspace*{5mm} \textbf{end if}\\
	\hspace*{5mm} \textbf{return}\\
	\vspace*{1em}
	\\
	\underline{$\mathsf{Dec}_{Q_{\mathcal{A}}}\left(\right)$}
	\vspace*{0.75em}
	\\				
	\hspace*{5mm} Apply $\mathbf{U}_{\mathcal{D}_K \left(\cdot\right)}$ to $Q_{\mathcal{A}}$\\
	\hspace*{5mm} \textbf{return}\\
\end{tabular}
\caption{The RoP-qsCCA confidentiality notion}
\label{fig:RoPqsCCA}
\end{figure}

\begin{definition}[\textbf{RoP-qsCCA}]
Let $\mathcal{S}\mathcal{E}=\left(\mathcal{K}, \mathcal{E}, \mathcal{D}\right)$
be a symmetric encryption scheme.  Define experiment
$\textbf{Exp}_{\mathcal{S}\mathcal{E}}^{rop-qscca-b}\left(\mathcal{A}\right)$
for a quantum adversary $\mathcal{A}$ and a bit $b$ as in
Figure~\ref{fig:RoPqsCCA}.  In the experiment, the adversary $\mathcal{A}$ is
given quantum superposition access to a real-or-permutation encryption oracle
$\mathsf{RoP}_{Q_{\mathcal{A}}}\left(\right)$ as well as a decryption oracle,
$\mathsf{Dec}_{Q_{\mathcal{A}}}\left(\right)$.

Now denote by $\rho^e_i$ the state of the register $Q_\A$ directly after the
$i$-th encryption query, and by $\rho^d_j$ the state of the register
directly before the $j$-th decryption query. For any (classical) ciphertext $c$ that occurs
as the result of an encryption query, i.e., for any ciphertext $c$ such that
$\langle c|\rho^e_i|c\rangle\ne0$ for some $i$, we require 
$\langle c|\rho^d_j|c\rangle=0$ for all $j$.
	
The adversary's goal is to output a bit $b'$ as its guess of the challenge bit
$b$, and the experiment returns $b'$ as well.  The advantage of $\mathcal{A}$ is given by
\[
\textbf{Adv}^{rop-qscca}_{\mathcal{S}\mathcal{E}}\left(\mathcal{A}\right) = 
{\rm Pr}\left[\textbf{Exp}_{\mathcal{S}\mathcal{E}}^{rop-qscca-1}\left(\mathcal{A}\right) = 1 \right] - 
{\rm Pr}\left[\textbf{Exp}_{\mathcal{S}\mathcal{E}}^{rop-qscca-0}\left(\mathcal{A}\right) = 1 \right] \;.
\]
This advantage refers to a specific quantum adversary using resources as
defined in Section~\ref{sec:model}.\qed
\end{definition}

The restriction on the adversary's decryption queries could be made less
restrictive by replacing the condition $\langle c|\rho^d_j|c\rangle=0$ by
$\langle c|\rho^d_j|c\rangle<\delta$ for some $\delta>0$. The complication
entailed by this did not, however, seem justified as, due to the randomization,
the ciphertext space is much larger than the message space. It is worth
pointing out it cannot be checked even in principle whether the adversary
honors the restriction on its decryption queries. This is because, in a
superposition query, the oracle (or experiment) can have no information on
what messages or ciphertexts are submitted as part of the query.

We now show that the above indistinguishability notion can be achieved by
the following standard Encrypt-then-MAC construction \cite{Bellare2000}:\\

\noindent{\bf Construction 2:} Let $\SE = \left(\mathcal{E},
\mathcal{D}\right)$ be a symmetric encryption scheme and let $F$ be a QPRF. 
The following construction
defines a symmetric encryption scheme 
$\mathcal{S}\mathcal{E'} = \left(\mathcal{E'}, \mathcal{D'}\right)$:
\begin{eqnarray*}
\mathcal{E'}\left(\left(K_1, K_2\right), m\right) :\, && c \leftarrow
\mathcal{E}_{K_1}\left(m\right), \, \tau \leftarrow F_{K_2}\left(c\right)\\ 
&& \textbf{\em output} \; \left( c, \tau \right) \\ 
\mathcal{D'}\left(\left(K_1,K_2\right), c, \tau \right):\, && \tau' \leftarrow F_{K_2}\left(c\right), \, m
\leftarrow \mathcal{D}_{K_1}\left(c\right)\\ 
&& \textbf{\em if} \; \tau = \tau',\, \textbf{\em output} \; \left( m\right)\\ 
&& \textbf{\em otherwise, output} \;
\bot
\end{eqnarray*}

The symbol $\bot$ denotes some fixed string that is outside the message space.
The decryption returns $\bot$ if the tag is invalid. A quantum decryption oracle
is described by a unitary operator
$\mathbf{V}$ acting on a register state $\left|x,c,\tau\right\rangle$ as follows:
\begin{equation}   \label{eq:bot}
\mathbf{V} \left|x,c,\tau\right\rangle = 
\left\{
\begin{array}{ll}
\left|x \oplus \mathcal{D}_{K_1}\left(c\right),c,\tau \right\rangle & \mbox{ if } F_{K_2}\left(c\right) = \tau\;,\\
\left|x \oplus \bot,c,\tau \right\rangle & \mbox{ otherwise.}
  \end{array} \right. 
\end{equation}

The following theorem establishes that, if the above construction can be broken
by a specific quantum adversary, then there exist quantum adversaries using
similar resources that break the underlying QPRF or the RoP-qsCPA
security of the underlying scheme $\SE$. But as we saw in
Sections~\ref{sec:qprf} and~\ref{sec:cpa}, a QPRF based on a suitably chosen
block cipher is currently thought to be secure against quantum attacks, and
RoP-qsCPA security is achievable.

\begin{theorem}[RoP-qsCCA security is achievable]
\label{thm:qsCCA}
Consider the scheme $\mathcal{S}\mathcal{E'}$ in Construction~2 based on a QPRF
$F$ and an encryption scheme $\SE$.  Assume $\A$ is a quantum adversary
attacking $\SE'$ in the RoP-qsCCA sense,
making at most $q_e$ encryption and $q_d$ decryption queries to the oracle, and
having advantage
\[	
\textbf{\em Adv}_{\mathcal{S}\mathcal{E'}}^{rop-qscca}\left(\mathcal{A}\right) \ge \epsilon \;.
\]
Then there exist quantum adversaries $\mathcal{B}$ and $\mathcal{J}$ attacking
$\mathcal{S}\mathcal{E}$ and $F$ respectively, as follows.  $\mathcal{B}$ makes
at most $q_e$ encryption oracle queries.  $\mathcal{J}$
makes at most $q_d$ oracle queries.  The advantages satisfy
\[
\textbf{\em Adv}_{\mathcal{S}\mathcal{E}}^{rop-qscpa}\left(\mathcal{B}\right) +
2 \cdot \textbf{\em Adv}_F^{qprf}\left(\mathcal{J}\right) \ge \epsilon -
2\left(1+2 q_d^2\right) 2^{-n_\tau/4} \;,
\]
where $n_{\tau}$ is the length of the tag $\tau$ as defined in Construction~2.
\end{theorem}

\proof As in the proof of Theorem~\ref{thm:qsCPA}, we first modify
Construction~2 by replacing $F$ with a true random function $f$. We then show,
roughly, (i) that a quantum adversary that can distinguish between these two
constructions can break the QPRF security of $F$, and (ii) that a quantum
adversary that breaks the modified construction can break the CPA security of
the underlying scheme $\SE$. 

We denote the modified scheme by $\widetilde{\SE} = (\widetilde{\E},\widetilde{\D})$:
\begin{eqnarray*}
\widetilde{\E}\left(K_1,m\right) :\, && c \leftarrow \E_{K_1}\left(m\right), \, \tau \leftarrow f(c)\\
&& \textbf{output} \; \left( c, \tau \right)\\ 
\widetilde{\D}\left(K_1,c,\tau \right):\, && \tau' \leftarrow f(c), \, m \leftarrow \mathcal{D}_{K_1}\left(c\right)\\
&& \textbf{if} \; \tau = \tau', \, \textbf{output} \; \left( m\right)\\
&& \textbf{otherwise, output} \; \bot
\end{eqnarray*}
Here, $f$ is a function chosen uniformly at random. 

Now let $\A$ be a quantum adversary attacking the CCA security of $\SE'$. 
A standard argument \cite{Katz2007} then leads to the construction of a quantum
adversary $\J$ attacking the QPRF $F$, making $q_d$ oracles queries and having advantage 
\begin{equation}
\label{eq:j}
\textbf{Adv}^{qprf}_{F}(\J)\ge
\frac12\textbf{Adv}_{\SE'}^{rop-qscca}(\A)-\frac12\textbf{Adv}_{\widetilde{\SE}}^{rop-qscca}(\A) \;,
\end{equation}
where the right hand side is one-half the difference between the advantages of
$\A$ in the experiments $\textbf{Exp}_{\SE'}^{rop-qscca-b}(\A)$ and
$\textbf{Exp}_{\widetilde\SE}^{rop-qscca-b}(\A)$ respectively.

To represent the actions of the oracles on the register $Q_\A$ of the quantum
adversary $\A$, it is convenient to split $Q_\A$ in three sections, as follows.
In the case of $\widetilde\SE$, the action of the encryption oracle is then
given by
\begin{equation}
\left|m,x,y\right\rangle\longrightarrow |m, x\oplus c,y\oplus\tau\rangle\;,
\end{equation}
where
\begin{equation}
c = \left\{
\begin{array}{ll}
\E_{K_1}(m) & \mbox{ if } b=1 \;,\\
\E_{K_1}(\Pi(m)) & \mbox{ if } b=0 \;,
\end{array} \right.
\end{equation}
and  $\tau=f(c)$, and the action of the decryption oracle is given by
\begin{equation}
\left|x,c,\tau\right\rangle \longrightarrow
\left|x\oplus\widetilde{\D}_{K_1}(c, \tau), c,\tau\right\rangle \;.
\end{equation}

We can use a quantum adversary $\mathcal{A}$ attacking the RoP-qsCCA security
of $\widetilde{\mathcal{S}\mathcal{E}}$ to construct a quantum adversary
$\mathcal{B}$ attacking the RoP-qsCPA security of $\mathcal{S}\mathcal{E}$
using $q_e$ encryption oracles queries. The quantum adversary $\mathcal{B}$
runs $\mathcal{A}$, and uses its oracles to provide a simulation of
$\mathcal{A}$'s oracles in the RoP-qsCCA experiment. $\mathcal{B}$ simulates
$\mathcal{A}$'s decryption queries by answering all of them with $\bot$.
We need to link the advantages of $\A$ and $\B$.

The main difficulty here is to derive a bound on the additional advantage of
$\A$ due to its ability to make decryption queries. In a classical (non-quantum)
setting, since $f$ is a true random function, the probability that an adversary
forges a valid tag for a ciphertext would be $q_d/2^{n_\tau}$.  A classical
adversary therefore would get $\bot$ almost every time in response to its
decryption queries, which means that the decryption oracle would be essentially
useless to the classical adversary. But because our adversary $\A$ is able to
make superposition queries, we have to work harder.

As in the classical proof, we will derive the required bound by considering a modified decryption oracle
which always returns $\bot$ in response to the quantum adversary's decryption queries,
irrespective of the values of $c$ and $\tau$. We will refer to the experiment
with the modified oracle as {\it scenario Q0}, and to the original experiment
as {\it scenario Q1}.

In scenario Q1, the decryption oracle returns $\bot$ only if the tag is invalid.
Similary to Eq.~(\ref{eq:bot}), the action of the decryption oracle is thus described by the unitary operator
$\mathbf{V}$ acting on a register state $\left|x,c,\tau\right\rangle$ as follows:
\begin{equation}
\mathbf{V} \left|x,c,\tau\right\rangle = 
\left\{
\begin{array}{ll}
\left|x \oplus \mathcal{D}_{K_1}\left(c\right),c,\tau \right\rangle & \mbox{ if } f\left(c\right) = \tau\;,\\
\left|x \oplus \bot,c,\tau \right\rangle & \mbox{ otherwise.}
  \end{array} \right. 
\end{equation}
The time evolution of the quantum adversary can then be written as
\begin{equation}
\label{eq:v}
\mathbf{U}_{q_d} \mathbf{V} \ldots \mathbf{U}_2 \mathbf{V} \mathbf{U}_1 \mathbf{V} \mathbf{U}_0 \left|s\right\rangle\;,
\end{equation}
followed by a binary measurement whose outcome is the guess $b'$.  The input
state $|s\rangle$ is the result of some initialisation.  The unitary operators
$\mathbf{U}_i$ describe the evolution of the adversary between decryption
queries and include the actions of the encryption oracle.
The probability of outcome $b'$ depends on the bit $b$ in the experiment. 
We will denote the probability that the outcome in scenario Q1 is $b'=1$ for
the two cases $b=0$ and $b=1$ by ${\rm Pr}^{Q1-0}(b'=1)$ and 
${\rm Pr}^{Q1-1}(b'=1)$, respectively. We have then
\begin{equation}
\label{eq:a}
\textbf{Adv}_{\widetilde{\SE}}^{rop-qscca}(\A)={\rm Adv}^{Q1} = {\rm Pr}^{Q1-1}(b'=1) - {\rm Pr}^{Q1-0}(b'=1) \;,
\end{equation}
where we have introduced the notation ${\rm Adv}^{Q1}$ for convenience.

The only difference in scenario Q0 is that the decryption oracle always returns
$\bot$. We denote the action of the decryption oracle in this case by 
$\tilde{\mathbf{V}}$, which acts like this:
\begin{equation}
\tilde{\mathbf{V}} \left|x,c,\tau\right\rangle = \left|x \oplus \bot,c,\tau \right\rangle \;.
\end{equation}
The time evolution of the quantum adversary in scenario Q0 is then given by 
\begin{equation}
\label{eq:vtilde}
\mathbf{U}_{q_d} \tilde{\mathbf{V}} \ldots \mathbf{U}_2 \tilde{\mathbf{V}} \mathbf{U}_1 \tilde{\mathbf{V}} \mathbf{U}_0 \left|s\right\rangle\;,
\end{equation}
again followed by a binary measurement whose outcome is the guess $b'$.  We
will denote the probability that the outcome in scenario Q0 is $b'=1$ for the
two cases $b=0$ and $b=1$ by ${\rm Pr}^{Q0-0}(b'=1)$ and ${\rm
  Pr}^{Q0-1}(b'=1)$, respectively. Since in scenario Q0 the decryption oracle
always return $\bot$, it is not useful for the quantum adversary
$\mathcal{A}$. This leads to a bound on the CPA advantage of the adversary $\B$:
\begin{equation}
\label{eq:b}
\textbf{Adv}^{rop-qscpa}_{\SE}(\B) \ge 
{\rm Adv}^{Q0} = {\rm Pr}^{Q0-1}(b'=1) - {\rm Pr}^{Q0-0}(b'=1) \;,
\end{equation}
where we have introduced the notation ${\rm Adv}^{Q0}$ again for convenience.

The key relation between the probabilities for scenarios Q0 and Q1 is provided
by the following {\bf claim:} 
\begin{equation}
\label{eq:claim}
{\rm Pr}^{Q1-b}(b'=1) \le  {\rm Pr}^{Q0-b}(b'=1) + (1 + 2 n_d^2)\,2^{-n_\tau/4} \;,
\end{equation}
independently of the value of the bit $b$. 
From inequality~(\ref{eq:claim}), together with the unproblematic assumption that 
${\rm Pr}^{Q0-b}(b'=1) \le {\rm Pr}^{Q1-b}(b'=1)$, 
we can deduce
\begin{eqnarray}
&& |{\rm Adv}^{Q1}-{\rm Adv}^{Q0}| \nonumber\\  &&=
|{\rm Pr}^{Q1-1}(b'=1) - {\rm Pr}^{Q1-0}(b'=1) - \big({\rm Pr}^{Q0-1}(b'=1) -{\rm
Pr}^{Q0-0}(b'=1)\big)| \nonumber\\  &&=
|{\rm Pr}^{Q1-1}(b'=1) - {\rm Pr}^{Q0-1}(b'=1) - \big({\rm Pr}^{Q1-0}(b'=1) - {\rm
Pr}^{Q0-0}(b'=1)\big)| 
\nonumber\\  &&\le
|{\rm Pr}^{Q1-1}(b'=1) - {\rm Pr}^{Q0-1}(b'=1)| + |{\rm Pr}^{Q1-0}(b'=1) - {\rm
Pr}^{Q0-0}(b'=1)|  
\nonumber\\  &&\le 2(1 + 2 n_d^2)\,2^{-n_\tau/4} \;.
\end{eqnarray}
Together with Eqs.~(\ref{eq:j}), (\ref{eq:a}) and ~(\ref{eq:b}), this implies
\begin{equation}
\textbf{\em Adv}_{\mathcal{S}\mathcal{E}}^{rop-qscpa}\left(\mathcal{B}\right) +
2 \cdot \textbf{\em Adv}_F^{qprf}\left(\mathcal{J}\right) \ge \textbf{Adv}_{\SE'}^{rop-qscca}(\A) -
2\left(1+2 q_d^2\right) 2^{-n_\tau/4} \;,
\end{equation}
which proves the theorem.

All that remains to be done is therefore to prove the claim~(\ref{eq:claim}). 
We start by examining the expressions~(\ref{eq:v}) and~(\ref{eq:vtilde}). 
Since, in general, the unitaries $\mathbf{U}_i$ entangle the adversary's
quantum register with its internal registers, one cannot assume that the
quantum register is in a pure state during decryption queries.  Denote by
$\mathcal{C}$ the set of all (classical) ciphertexts $c$.  For any ciphertext $c\in
\mathcal{C}$, define the projector
\begin{equation}
	\mathsf{Proj}_c = \sum_{m,\tau} \left|m,c,\tau \right\rangle \!\! \left\langle m,c,\tau\right| = I\otimes \left|c\right\rangle \!\! \left\langle c\right| \otimes I \;.
\end{equation}

Denote by $\rho^e_i$ the state of the quantum register after the $i$-th encryption query in scenario Q1. 
Let $\mathcal{C'}$ be the set of all ciphertexts that do not result from any encryption query.
That is, $\mathcal{C'}$ is the set of ciphertexts that have zero weight in all
encryption queries, i.e., 
\begin{equation}
  \mathcal{C'}=\left\{c\in\mathcal{C}\,:\,{\rm Tr}\left(\mathsf{Proj}_c\rho^e_i\right)=0 \;, i=1,\ldots,q_e\right\} \;,
\end{equation}
where Tr denotes the trace. We can now define the set $\mathcal{C}_{{\rm valid}}$ as the set of pairs
$(c,f(c))$ that do not result from any encryption query,
\begin{equation}
\mathcal{C}_{{\rm valid}} = \left\{ \left(c,f(c)\right) \,:\, c\in \mathcal{C'} \right\} \;.
\end{equation}
Since
\begin{equation}
\left|\mathcal{C}_{{\rm valid}}\right| = 2^{-n_\tau} \left|\mathcal{C'} \times \left\{0,1\right\}^{n_\tau}\right| \;,
\end{equation}
trying to guess $\tau = f\left(c\right)$ given a ciphertext $c\in \mathcal{C'}$
leads to a valid pair with very small probability.  The results of the $q_e$
encryption queries contain no information about the set $\mathcal{C}_{{\rm valid}}$.

Now let $\rho^d_i$ be the state of the quantum register before the $i$-th decryption query in scenario Q1, and define
\begin{equation}
\mathsf{Proj}_{{\rm valid}} =  \sum_{(c,\tau)\in \mathcal{C}_{{\rm valid}}} |c,\tau \rangle \!\! \langle c,\tau|  \;.
\end{equation}
We can now define $\mathsf{W}_{{\rm val},i}$ as the total weight of terms
belonging to $\mathcal{C}_{{\rm valid}}$ in the $i$-th decryption query
($i=1,\ldots,q_d$),
\begin{equation}
	\mathsf{W}_{{\rm val},i} = {\rm Tr}\left(\rho^d_i \, \mathsf{Proj}_{{\rm valid}}\right) \;.
\end{equation}
This can be re-expressed as follows. Let $\left|\psi^d_i\right\rangle$ be the
state of the totality of the adversary's quantum registers immediately before
the $i$-th decryption query in scenario Q1,
\begin{equation}
\left|\psi^d_i\right\rangle 
= \mathbf{U}_{i-1} \mathbf{V}\ldots \mathbf{U}_1 \mathbf{V}\mathbf{U}_0 \left|s\right\rangle \;,
\end{equation}
which can be expanded in the form
\begin{equation}
\left|\psi^d_i\right\rangle = \sum_{j,m,c,\tau} \lambda_{j,m,c,\tau} \left|j,m,c,\tau \right\rangle \;,
\end{equation}
where $j$ labels the computational basis states of all internal registers
(i.e., all registers in addition to the register $Q_\A$).  We then have
\begin{equation} 
\label{eq:Wval}
\mathsf{W}_{{\rm val},i} = \left\langle \psi^d_i \right| \mathsf{Proj}_{{\rm
    valid}} \left|\psi^d_i \right\rangle = \sum_{j,m,c}
\left|\lambda_{j,m,c,f\left(c\right)}\right|^2 = 1 - \sum_{j,m,c,\tau \ne
  f\left(c\right)} \left|\lambda_{j,m,c,\tau}\right|^2\;.
\end{equation}
The probability that a direct measurement after the $i$-th decryption query
gives a string $\left(c,\tau\right)\in \mathcal{C}_{{\rm valid}}$ is then given
by the expectation value $\mathsf{E}\left(\mathsf{W}_{{\rm val},i}\right)$.

Now the optimal way of searching for a string $\left(c,\tau\right)\in
\mathcal{C}_{{\rm valid}}$ is Grover's algorithm
\cite{Grover1996,Nielsen2000}.
As long as $i$ is less than
the minimum number of queries required for Grover's algorithm to succeed with
certainty (which is approximately $\frac{\pi}{4}\sqrt{2^{n_\tau}}$), the best
probability with which any quantum algorithm can find a string
$\left(c,\tau\right)\in \mathcal{C}_{{\rm valid}}$ using $i$ queries is exactly
the probability ${\rm Pr}_{{\rm Grover}}$ achieved by running Grover's
algorithm with $i$ queries \cite{Zalka1997,Ambainis2005}.  That probability is
equal to ${\rm Pr}_{{\rm
    Grover}}=\sin^2\left(\left(i+\frac{1}{2}\right)\theta\right)$, where
$\sin\frac{\theta}{2}=\sqrt{2^{-n_\tau}}$ \cite{Nielsen2000}.  To a very good
approximation,
\begin{equation}
	{\rm Pr}_{{\rm Grover}} = 4i^2\,2^{-n_\tau} \;.
\end{equation}

By measuring the quantum register after the $i$-th query and then stopping, the
quantum adversary can find a string $\left(c,\tau\right) \in \mathcal{C}_{{\rm
    valid}}$ with probability $\mathsf{E}\left(\mathsf{W}_{{\rm
    val},i}\right)$.  Therefore we must have
\begin{equation}
	\mathsf{E}\left(\mathsf{W}_{{\rm val},i}\right) \le 4i^2\,2^{-n_\tau}
\end{equation}
for $i=1,\ldots,q_d$.  What we actually need is a bound on the probabilities
for $\sqrt{\mathsf{W}_{{\rm val},i}}$.  For any random variable $X \ge 0$, we
have $\mathsf{Var}(\sqrt{X})\ge 0$  and thus $\mathsf{E}(\sqrt{X})\le
\sqrt{\mathsf{E}(X)}$.  Hence,
\begin{equation}
	\mathsf{E}\left(\sqrt{\mathsf{W}_{{\rm val},i}}\right) \le 2i\,2^{-n_\tau/2} \;.
\end{equation}

Now we want to compare the probability of outputting the guess $b'=1$ in
scenario Q0 and the probability of outputting the guess $b'=1$ in scenario Q1.
Let $\left|\psi^d_i\right\rangle$ denote the state immediately before the
$i$-th decryption query in scenario Q1 as before and, similarly, let
\begin{equation}
|\widetilde\psi^d_i\rangle
=\mathbf{U}_{i-1}\tilde{\mathbf{V}}\ldots\mathbf{U}_1\tilde{\mathbf{V}}\mathbf{U}_0 \left|s\right\rangle 
\end{equation}
denote the state immediately before the $i$-th decryption query in scenario Q0. 
Let us first compare the action of ${\mathbf{V}}$ and $\tilde{\mathbf{V}}$ on
the state $\left|\psi^d_i\right\rangle$ for some $i$:
\begin{eqnarray}
\mathbf{V}|\psi^d_i\rangle 
&=& \sum_{j,m,c,\tau\ne f(c)} \lambda_{j,m,c,\tau}\mathbf{V}|j,m,c,\tau\rangle 
+ \sum_{j,m,c,\tau=f(c)} \lambda_{j,m,c,\tau} \mathbf{V}|j,m,c,\tau\rangle  \nonumber \\
&=& \sum_{j,m,c,\tau\ne f(c)} \lambda_{j,m,c,\tau} |j,m \oplus \bot,c,\tau\rangle  \nonumber \\
&& \hspace*{5mm} + \sum_{j,m,c,\tau=f(c)} \lambda_{j,m,c,\tau} |j,m \oplus \mathcal{D}_{K_1}(c),c,\tau\rangle \;,
\end{eqnarray}
and
\begin{eqnarray}
\tilde{\mathbf{V}}\left|\psi^d_i\right\rangle  
&=& \sum_{j,m,c,\tau\ne f(c)} \lambda_{j,m,c,\tau} \tilde{\mathbf{V}}|j,m,c,\tau\rangle 
+ \sum_{j,m,c,\tau=f(c)} \lambda_{j,m,c,\tau} \tilde{\mathbf{V}}|j,m,c,\tau\rangle  \nonumber \\
&=& \sum_{j,m,c,\tau\ne f(c)} \lambda_{j,m,c,\tau} |j,m \oplus \bot,c,\tau\rangle  \nonumber \\
&& \hspace*{5mm} + \sum_{j,m,c,\tau=f(c)} \lambda_{j,m,c,\tau}|j,m \oplus \bot,c,\tau\rangle \;.
\end{eqnarray}
Putting these together and using Eq.~(\ref{eq:Wval}) twice, we get the
following for the fidelity of these two states:
\begin{eqnarray}
&& \left| \left\langle \psi_i^d\right|\tilde{\mathbf{V}}^{\dagger} \mathbf{V}
\left|\psi_i^d \right\rangle \right| \nonumber\\
&& = \left|\sum_{j,m,c,\tau\ne f\left(c\right)} \left|\lambda_{j,m,c,\tau}\right|^2 
 + \sum_{j,m,m',c} \lambda^*_{j,m',c,f\left(c\right)} \lambda_{j,m,c,f\left(c\right)} \left\langle m' \oplus \bot | m \oplus \mathcal{D}_{K_1}\left(c\right) \right\rangle \right| \nonumber \\
&&= \left|\sum_{j,m,c,\tau\ne f\left(c\right)} \left|\lambda_{j,m,c,\tau}\right|^2 + \sum_{j,m,c} \lambda^*_{j,m \oplus \mathcal{D}_{K_1}\left(c\right) \oplus \bot,c,f\left(c\right)} \lambda_{j,m,c,f\left(c\right)} \right| \nonumber \\
&&\ge \left|\sum_{j,m,c,\tau\ne f\left(c\right)} \left|\lambda_{j,m,c,\tau}\right|^2 \right| - 
	\left|\sum_{j,m,c} \lambda^*_{j,m \oplus \mathcal{D}_{K_1}\left(c\right) \oplus \bot,c,f\left(c\right)} \lambda_{j,m,c,f\left(c\right)}\right| \nonumber \\
	&&= 1-\mathsf{W}_{{\rm val},i} - \left|\sum_{j,m,c} \lambda^*_{j,m \oplus \mathcal{D}_{K_1}\left(c\right) \oplus \bot,c,f\left(c\right)} \lambda_{j,m,c,f\left(c\right)}\right| \nonumber \\
	&&\ge 1-\mathsf{W}_{{\rm val},i} - \sqrt{\sum_{j,m,c}\left|\lambda_{j,m \oplus \mathcal{D}_{K_1}\left(c\right) \oplus \bot,c,f\left(c\right)}\right|^2}\sqrt{\sum_{j,m,c}\left|\lambda_{j,m,c,f\left(c\right)}\right|^2} \nonumber \\
	&&= 1-\mathsf{W}_{{\rm val},i} - \sqrt{\mathsf{W}_{{\rm val},i}}\,\sqrt{\mathsf{W}_{{\rm val},i}} \nonumber \\
	&&= 1-2\,\mathsf{W}_{{\rm val},i} \;.
\end{eqnarray}
This implies that the trace distance \cite{Nielsen2000} of these two states is bounded as
\begin{equation}
\mathsf{D}\left(\mathbf{V}\left|\psi^d_i \right\rangle, \tilde{\mathbf{V}}\left|\psi^d_i \right\rangle \right) \le \sqrt{1-\left(1-2\,\mathsf{W}_{{\rm val},i}\right)^2} \le 2\sqrt{\mathsf{W}_{{\rm val},i}} \;.
\end{equation}

Before the first decryption query, the states of the adversary in both scenarios
Q0 and Q1 are identical, 
\begin{equation}
\left|\psi_1^d \right\rangle =
\mathbf{U}_0\left|s\right\rangle \;.
\end{equation}
Before the second decryption query, the states are 
\begin{equation}
|\widetilde{\psi}_2^d \rangle
 = \mathbf{U}_1
\tilde{\mathbf{V}} |\psi_1^d \rangle
 \;\mbox{ and }\;
|\psi_2^d\rangle 
= \mathbf{U}_1 \mathbf{V} |\psi_1^d \rangle \;,
\end{equation}
respectively.  Therefore, for the trace distance we have
\begin{equation}
	\mathsf{D}\left(|\psi_2^d \rangle,|\widetilde{\psi}_2^d \rangle \right) =
	 \mathsf{D}\left(\mathbf{V}|\psi^d_1\rangle,
         \tilde{\mathbf{V}}|\psi^d_1\rangle \right) 
\le 2\sqrt{\mathsf{W}_{{\rm val},1}} \;.
\end{equation}
For arbitrary $i>0$, the triangle inequality gives us
\begin{eqnarray}
\mathsf{D}\left(|\psi^d_{i+1} \rangle,|\widetilde{\psi}^d_{i+1} \rangle \right)
&=& \mathsf{D}\left(\mathbf{U}_i\mathbf{V}|\psi^d_i\rangle,\mathbf{U}_i\tilde{\mathbf{V}}|\widetilde{\psi}^d_i\rangle \right) \nonumber  \\ 
&=& \mathsf{D}\left(\mathbf{V}|\psi^d_i\rangle,\tilde{\mathbf{V}}|\widetilde{\psi}^d_i\rangle \right)  \nonumber \\
&\le& \mathsf{D}\left(\mathbf{V}|\psi^d_i\rangle,\tilde{\mathbf{V}}|\psi^d_i\rangle \right) 
+ \mathsf{D}\left(\tilde{\mathbf{V}}|\psi^d_i\rangle,\tilde{\mathbf{V}}|\widetilde{\psi}^d_i\rangle \right)   \nonumber  \\
&=& \mathsf{D}\left(\mathbf{V}|\psi^d_i\rangle,\tilde{\mathbf{V}}|\psi^d_i\rangle \right) 
+ \mathsf{D}\left(|\psi^d_i\rangle,|\widetilde{\psi}^d_i\rangle \right)   \nonumber  \\
&\le& 2\sqrt{\mathsf{W}_{{\rm val},i}} + \mathsf{D}\left(|\psi^d_i\rangle,|\widetilde{\psi}^d_i\rangle \right) \;. 
\end{eqnarray}
By induction, it follows that
\begin{equation}
\label{eq:induction}
\mathsf{D}\left(\left|\psi^d_{q_d} \right\rangle,\left|\widetilde{\psi}^d_{q_d} \right\rangle \right) \le 2 \sum_{i=1}^{q_d-1}\sqrt{\mathsf{W}_{{\rm val},i}} \;.
\end{equation}

This implies that, for any measurement, the probabilities for $b'=1$ in both
scenarios can not differ by more than the right-hand side of
Eq.~(\ref{eq:induction}).
Now the expectation of that quantity is
\begin{eqnarray}
\mathsf{E}\left(2 \sum_{i=1}^{q_d-1}\sqrt{\mathsf{W}_{{\rm val},i}}\right) & =& 2 \sum_{i=1}^{q_d-1} \mathsf{E}\left(\sqrt{\mathsf{W}_{{\rm val},i}}\right) \nonumber \\
	& \le& 2^{-n_\tau/2}4\sum_{i=1}^{q_d-1} i \nonumber \\
	& \le& 2 q_d^2\,2^{-n_\tau/2} \;.
\end{eqnarray}
Using the Markov inequality this implies, for any $\xi>0$, 
\begin{eqnarray}
{\rm Pr} \left(2\sum_{i=1}^{q_d-1}\sqrt{\mathsf{W}_{{\rm val},i}} \ge \xi\right) 
& \le& \frac{1}{\xi}\, \mathsf{E}\left(2 \sum_{i=1}^{q_d-1}\sqrt{\mathsf{W}_{{\rm val},i}}\right) \nonumber \\
& \le& \frac{2}{\xi} q_d^2\,2^{-n_\tau/2} \;.
\end{eqnarray}
That is, with probability at least $1-\frac{2}{\xi} q_d^2\,2^{-n_\tau/2}$, we have that
\begin{equation}
{\rm Pr}^{Q1-b}(b'=1) \le  {\rm Prob}^{Q0-b}(b'=1) + \xi\;,
\end{equation}
irrespectively of the value of the bit $b$. It follows that 
\begin{equation}
{\rm Pr}^{Q1-b}(b'=1) \le  {\rm Pr}^{Q0-b}(b'=1) + \xi + \frac2\xi n_d^2\,2^{-n_\tau/2} \;.
\end{equation}
We can now choose $\xi$ so that this has the most convenient form. One possibility is
$\xi=2^{-n_\tau/4}$, which leads to 
\begin{equation}
{\rm Pr}^{Q1-b}(b'=1) \le  {\rm Pr}^{Q0-b}(b'=1) + (1 + 2 n_d^2)\,2^{-n_\tau/4} \;,
\end{equation}
which establishes the claim (\ref{eq:claim}). This completes the proof of
the theorem. \qed

%--------------------------------------------------------------------

\end{document}